\documentclass[a4paper,superscriptaddress,nofootinbib,12pt]{revtex4}
\usepackage{eurosym}
\usepackage{amsfonts}
\usepackage{amsmath}
\usepackage{amsthm}
\usepackage{bbm}
\usepackage{amssymb}
\usepackage{mathrsfs}
\usepackage{stmaryrd}
\usepackage{dsfont}
\usepackage{wasysym}
\usepackage[a4paper]{geometry}
\begin{document}

\title{Heisenberg doubles for Snyder type models}
\author{Stjepan Meljanac}
\email{meljanac@irb.hr}
\affiliation{Division of Theoretical Physics, Rudj{}er Bo\v{s}kovi\'c Institute, Bijeni%
\v{c}ka~c.54, HR-10002~Zagreb, Croatia}
\author{Anna Pacho{\l }}
\email{a.pachol@qmul.ac.uk}
\affiliation{Queen Mary, University of London, Mile End Rd., London E1 4NS, UK.\\
\phantom{P}}

\begin{abstract}
A Snyder model generated by the noncommutative coordinates and Lorentz
generators close a Lie algebra.  The application of the Heisenberg double construction is investigated for the Snyder coordinates and momenta generators. It leads to the phase space of the Snyder model.
Further, the extended
Snyder algebra is constructed by using the Lorentz algebra, in one dimension
higher. The dual pair of extended Snyder algebra and extended Snyder group
is then formulated. Two Heisenberg doubles are considered, one with the
conjugate tensorial momenta and another with the Lorentz matrices. Explicit
formulae for all Heisenberg doubles are given.
\end{abstract}

\maketitle

\section{Introduction}

Noncommutative coordinates and noncommutative spacetimes lead to a
modification of their corresponding relativistic symmetries, which are
described by quantum groups (Hopf algebras). Having a dual pair of Hopf
algebras (one describing the quantum symmetry group $A$ and its dual quantum
Lie algebra $A^*$), one can construct the so called Heisenberg double. The
Heisenberg double, although interesting from mathematical point of view, can
be also seen as a generalization of the quantum mechanics phase space
(Heisenberg algebra). Such constructions have been of interest especially
for the $\kappa$-Minkowski noncommutative spacetime and its deformed
relativistic symmetry $\kappa$-Poincar\'e quantum group \cite{KosMas}, \cite{GGKM}, \cite{Lukierski1}, \cite{Lukierski2}, \cite{TMPH} as well as for the 
$\theta$-deformation \cite{TMPH} and for the general Lie algebra type
noncommutative spaces \cite{Zoran}.

In the present paper we want to focus on the Heisenberg double construction
applied to a Snyder model. In the 1940s, Snyder proposed the model of
Lorentz invariant discrete spacetime \cite{Snyder40} as the first example of
the noncommutative spacetime.

The Snyder model has been attracting quite a lot of attention in the
literature \cite{BatMelj}-\cite{Tkachuk1}. Field theory on this space was
considered, for example in \cite{BatMelj}, \cite{Girelli}, \cite{field2}, its extension to a cosmological \cite%
{Battisti:2008du} and curved \cite{Mig1a}, \cite{Mig1b}
backgrounds was proposed, deformed Heisenberg uncertainty relations were
investigated \cite{PRD79}, different applications to quantum gravity
phenomenology have been considered as well, see, e.g. \cite{Battisti:2008du}%
, \cite{Mig2},\cite{Mig3}. Different Snyder phase spaces arising within the
projective geometry context were investigated in \cite{Balla}, \cite{Ballb}
and the $\kappa$-Snyder space with non associative star product was proposed
in \cite{kappaSnyder}.

In the Snyder model the coordinates do not commute and their commutation
relation is proportional to the Lorentz generators. For this reason
noncommutative coordinates by themselves do not close a Lie algebra and cannot
be equipped in the Hopf algebra structure. In this paper we investigate various ways
of extending the Snyder space so that we can define the Lie algebra
containing the Snyder coordinates. Then the Hopf algebra structure arises
naturally and the Heisenberg double construction may be attempted.

We start with the algebra generated by Snyder coordinates $\hat{x}_{i}$ and
Lorentz generators $M_{jk}$, so that it becomes a Lie algebra and then we
equip its universal enveloping algebra with the Hopf algebra structure, where the Lie algebra generators
have primitive coproducts. For the Heisenberg double construction we need the
dual Hopf algebra. Since the Hopf algebra of Snyder coordinates and Lorentz generators is associative, the dual Hopf algebra should be coassociative. However, we are interested in investigating the possibility of constructing the phase
space relations between Snyder coordinates and the commuting momenta $p_{i}$ with non-coassociative coproducts.
We try to follow the steps of the Heisenberg double construction and investigate its limitations in obtaining the commutation relations
between the Snyder coordinates and momenta generators. The cross-commutation relations obtained correspond to the known versions of the Snyder phase space
used in various applications, e.g. in \cite{Mig1a}, \cite{Mig1b}, \cite{Mig2}%
, \cite{Mig3}. 
Then the analogous cross-commutation relations are obtained for (non-coassociative) coproducts of momenta in
various realizations \cite{Snyder40}, \cite{BatMelj}. 

In the second part of the paper, to overcome the obstacles raised by non-coassociativity and to construct the full dual Hopf algebra and the full Heisenberg double for the Snyder model, we propose using the
extended noncommutative Snyder coordinates in the following sections.

We analyse the extended version of the Snyder
model, where Snyder coordinates are identified as $\hat{x}_{i}\sim \hat{x}%
_{iN}=M_{iN}\in so(1,N)/so(1,N-1)$ \cite{Girelli}, \cite{2007Snyder}. 
Thanks to this we are able to
construct two full Heisenberg doubles, firstly for the extended Snyder
algebra generated by tensorial coordinates $\hat{x}_{\mu \nu }$ with its
dual Hopf algebra generated by tensorial (conjugate) momenta $p_{\rho \sigma
}$. This way we find the Heisenberg double for the extended Snyder space
which may be considered as the extended Snyder phase space. Secondly, we
consider another Heisenberg double for the extended Snyder algebra with its
dual Hopf algebra of functions on a group ${\Lambda} _{\rho \sigma }$
(Lorentz matrices). We also present the Weyl realization for these Lorentz
matrices in terms of tensorial momenta. We finish the paper with brief conclusions.

\section{Issues with the Heisenberg double for the Snyder model}
\label{HD_Snyder} Snyder space is defined by the position operators $\hat{x}%
_{i}$ obeying the following commutation relations: 
\begin{equation}
\lbrack \hat{x}_{i},\hat{x}_{j}]=i\beta M_{ij}  \label{Snyder_orig1}
\end{equation}%
where $M_{ij}$ are the generators of the Lorentz algebra $so(1,N-1)$ and $%
\beta $ is the Snyder parameter of length square dimension (usually assumed
to be of order of Planck length $L_{p}^{2}$) that sets the scale of
noncommutativity (we use natural units $\hbar =c=1$). Note that here $%
i,j=0\dots ,N-1$.

In agreement with Snyder \cite{Snyder40} the symmetry of such
(noncommutative) space is described by the undeformed Lorentz algebra $%
so(1,N-1)$. This requires that the $M_{ij}$ generators satisfy the standard
commutation relations:
\begin{equation}
\lbrack M_{ij},M_{kl}]=i(\eta _{ik}M_{jl}-\eta _{il}M_{jk}+\eta
_{jl}M_{ik}-\eta _{jk}M_{il}).  \label{MM}
\end{equation}
We also have the cross-commutation relations between Lorentz generators and
Snyder coordinates: 
\begin{equation}
\lbrack M_{ij},\hat{x}_{k}]=i(\eta _{ik}\hat{x}_{j}-\eta _{jk}\hat{x}_{i}).
\label{Mx}
\end{equation}
Relations \eqref{Snyder_orig1}, \eqref{MM}, \eqref{Mx} constitute a Lie
algebra, which we embed into the associative universal enveloping algebra. We denote this universal enveloping algebra as the algebra $A$. 

We are interested in constructing the Heisenberg double corresponding to the noncommutative
Snyder space, therefore first we need to equip $A$ with the Hopf
algebra structure. It is enough to impose the primitive coalgebra structure on $\hat{x}_i$ and $M_{ij}$.
We investigate if it is possible to use the Heisenberg double construction to obtain the phase
space built up from the Snyder noncommutative coordinates $\hat{x}_{i}$ and the momenta $p_{i}$.
We consider the algebra $\tilde{A}$ generated by commuting momenta $p_{i}$ equipped in the non-coassociative coalgebra structure. We choose
the realization for the
coproducts of momenta which was proposed in \cite{BatMelj} and called the Snyder realization therein.

The defining relations of $\tilde{A}$ are:  
\begin{equation}
\left[ p_{i},p_{j}\right] =0
\end{equation}%
and the (non-coassociative) coalgebra structure is the following: 
\begin{eqnarray}
\Delta p_{i} &=&1\otimes p_{i}+\frac{1}{1-\beta p_{k}\otimes p^{k}}\left(
p_{i}\otimes 1-\frac{\beta }{1+\sqrt{1+\beta p^{2}}}p_{i}p_{j}\otimes p^{j}+%
\sqrt{1+\beta p^{2}}\otimes p_{i}\right) ,  \notag  \label{copp_Sn} \\
\epsilon (p_{i}) &=&0,\quad S(p_{i})=-p_{i},
\end{eqnarray}
where $p^{2}=\eta ^{ij}p_{i}p_{j}$ is Lorentz invariant and $i,j=0\dots ,N-1$. 

We note that $\tilde{A}$ is non-coassociative, therefore momenta cannot be the proper dual generators to Snyder coordinates which are the generators of the associative Hopf algebra $A$ (see, e.g. \cite{Klimyk}, Sec. 1.2.8).
Nevertheless, we can still propose the following
 duality relations $<\quad ,\quad >:\tilde{A}\times A\rightarrow 
\mathbb{C}$ (on the generators only)\footnote{The pairing we propose here is satisfied on the generators (in the first power) only and it may not be possible to extend it to the full algebra due to the non-coassociative nature of $\tilde{A}$. We also note that the proper definition of $\tilde{A}$ would require quasi-Hopf algebra framework \cite{Drinfeld_quasi}. Defining the Heisenberg double within the quasi-Hopf algebra setting has been proposed, for example in \cite{Panaite2}, \cite{Panaite} and would be worth investigating further.}: 
\begin{equation}
<p_{i},\hat{x}_{j}>=-i\eta _{ij},  \label{dual1}
\end{equation}
and
\begin{equation}
<p_{i},M_{jk}>=0.  \label{dual2}
\end{equation}
Then we consider the analogue of the left Hopf action $\triangleright $  
of $\tilde{A}$ on $A$ defined as 
\begin{eqnarray}\label{action}
p_{i}\triangleright \hat{x}_{j} &=&<p_{i},\hat{x}_{j_{\left( 2\right) }}>%
\hat{x}_{j_{\left( 1\right) }}=-i\eta _{ij}.
\end{eqnarray}%
(on the generators only) and use the usual cross product construction mimicking that of the
 Heisenberg double (we refer the
reader to the Appendix for the details of the Heisenberg double
construction in the Hopf algebra setting). The resulting cross-commutation relations are: 
\begin{equation}\label{p-x1}
\left[ p_{i},\hat{x}_{j}\right] =\hat{x}_{j}{}_{\left( 1\right) }<p_{i\left(
1\right) },\hat{x}_{j\left( 2\right) }>p_{i\left( 2\right) }-\hat{x}%
_{j}p_{i}=-i(\eta _{ij}+\beta p_{i}p_{j}).
\end{equation}%
We note that relations \eqref{p-x1} between momenta and Snyder coordinates
obtained here (although with the limitations discussed), are in agreement with the commutation
relations for the phase space of the Snyder model usually considered in the
literature, see, e.g., \cite{Mig1a}, \cite{Mig1b}, \cite{Mig2}, \cite{Mig3}. 
\subsection{Different realization for
coproducts of momenta}

In the previous section we have used the coalgebra sector for momenta in the
so-called Snyder realization \cite{Snyder40}, \cite{BatMelj}.
There exist more possible realizations for momenta's (non-coassociative)
coproducts, 
see e.g. \cite{BatMelj}, \cite{Maggiore_real}. Different realizations for coproducts can be related with each other by a change of basis in the momentum space.

However, there is a general way to write the (non-coassociative) coproduct
for momenta corresponding to the Snyder model, i.e. the realization proposed in \cite{BatMelj} and called 'general realization' therein \footnote{The general realization for the coproduct of momenta corresponds to the
general realization for the Snyder coordinates, see eq. (6), (7) in \cite%
{BatMelj}, where $\hat{x}_{i}\triangleright 1={x}_{i},\quad
M_{ij}\triangleright 1=0$.}.
The formula we recall below \cite%
{BatMelj} is calculated only up to the second order in the parameter $\beta $: 
\begin{eqnarray}
\Delta p_{i} &=&1\otimes p_{i}+p_{i}\otimes 1+  \notag  \label{copp_gen} \\
&&+\beta \left( \left( c-\frac{1}{2}\right) p_{i}\otimes p^{2}+\left( 2c-%
\frac{1}{2}\right) p_{i}p_{k}\otimes p^{k}+c\left( p^{2}\otimes
p_{i}+2p_{k}\otimes p^{k}p_{i}\right) \right) +O(\beta ^{2}), \\
\epsilon (p_{i}) &=&0,\quad S(p_{i})=-p_{i}.
\end{eqnarray}%
This formula describes the general non-coassociative \footnote{The difference between the left hand side and the right hand side of the coassociativity condition for the coproduct of momenta can be explicitly calculated and is as follows, in the first order in $\beta$: 
\begin{equation*}
\left( (id\otimes \Delta )\circ \Delta p_{i}\right) \,-\,\left( (\Delta
\otimes id)\circ \Delta p_{i}\right) =-\frac{1}{2}\beta \left( p_{i}\otimes
p_{k}\otimes p^{k}-p_{k}\otimes p_{i}\otimes p^{k}\right) +O\left( \beta
^{2}\right).
\end{equation*}} coproduct for Snyder momenta.
The choice of the parameter $c$ encodes different realizations.
For $c=\frac{1}{2}$ the coproduct \eqref{copp_gen} admits its finite form considered in the previous section.
For $c=0$ we get the realization investigated in \cite{Maggiore_real}.
 If $c=\frac{1}{6}$ we get the realization investigated in \cite{Weyl_real}, \cite{2003Melj}.

We can now calculate the cross-relations between momenta and Snyder coordinates corresponding to
the 'general realization' of the momenta coproducts for the Snyder model (the proposed duality relations on the generators and the left Hopf action proposed in the previous section remain unchanged \eqref{dual1},
 \eqref{dual2}, \eqref{action} for the linear power of the generators). It
results in the following cross-commutation relations between momenta and Snyder
coordinates: 
\begin{eqnarray}
\left[ p_{i},\hat{x}_{k}\right] &=&\hat{x}_{k}{}_{\left( 1\right)
}<p_{i\left( 1\right) },\hat{x}_{k\left( 2\right) }>p_{i\left( 2\right) }-%
\hat{x}_{k}p_{i}  \notag \\
&=&-i\eta _{ik}\left( 1+\beta \left( c-\frac{1}{2}\right) p^{2}\right)
-2ic\beta p_{k}p_{i}+O(\beta ^{2}).  \label{gen_real_p-x}
\end{eqnarray}
This reduces to (for the specific, above mentioned, choices of the parameter $c$):
\begin{itemize}
\item for $c=\frac{1}{2}$  \cite{BatMelj}  to:
\begin{equation}
\left[ p_{i},\hat{x}_{k}\right] =-i(\eta _{ik}+\beta p_{i}p_{k})
\end{equation}
cf. \eqref{p-x1};
\item for $c=0$ \cite{Maggiore_real} to:
\begin{equation}
\left[ p_{i},\hat{x}_{k}\right] =-i\eta _{ik}\left( 1-\frac{\beta }{2}%
p^{2}\right)+O(\beta ^{2});
\end{equation}
\item for $c=\frac{1}{6}$  \cite{Weyl_real}, \cite{2003Melj} to: 
\begin{equation}
\left[ p_{i},\hat{x}_{k}\right] =-i\eta _{ik}\left( 1-\frac{\beta }{3}%
p^{2}\right) -\frac{i}{3}\beta p_{k}p_{i}+O(\beta ^{2}).  \label{ph_spWeyl}
\end{equation}
\end{itemize}
It is worth to note that in the limit of $\beta \rightarrow 0$ the
coproducts for momenta \eqref{copp_Sn}, \eqref{copp_gen} reduce to $\Delta
p_{i}=1\otimes p_{j}+p_{i}\otimes 1$. And for $\beta = 0$ (classical case) there exists full duality between algebra $A$ generators $x_i$, $M_{ij}$ and group elements $p_i$, $\Lambda_{ij}$ (where $\Lambda_{ij}$ are matrix elements of Lorentz matrices, see e.g. \cite{KosMas}, \cite{GGKM}, \cite{Lukierski1}, \cite{Lukierski2}, \cite{TMPH}).

In this section we have focused on the Snyder space and we have tried to investigate the possibility of using the Heisenberg double procedure to obtain the phase space for this model. We have encountered the following issues. 
First, the noncommutative Snyder coordinates do not close the Lie algebra and only after extending the algebra by the Lorentz generators we have a Lie algebra which can be equipped in the Hopf algebra structure.
Second, the momenta corresponding to the Snyder model are non-coassociative hence the corresponding structure is not a Hopf algebra. Nevertheless, we try and propose the duality between the Snyder coordinates and momenta which is valid only on the generators (in the linear power). This allows us to mimic the Heisenberg double construction and leads to the cross-relations that are in agreement with the literature.
Additionally, we cannot define the dual elements to the Lorentz generators $M_{ij}$ for $\beta\neq 0$ (cf.  footnote 2). And the non-coassociative momenta do not allow for expanding the Lorentz algebra \eqref{MM} to the Poincar\'e algebra. 

The only way to construct the full dual Hopf algebra for the Snyder model and the full Heisenberg double is described in Sections \ref{unif} and \ref{extSn-HDs} and requires introducing the extended noncommutative coordinates $\hat{x}_{ij}$.

\section{Unified notation for the Snyder algebra: Extended Snyder model}\label{unif}

In the previous section, to consider a Hopf algebra related to the Snyder model, we have expanded the commutation relations between Snyder
coordinates \eqref{Snyder_orig1} by the Lorentz algebra \eqref{MM} and
included the cross-commutation relations \eqref{Mx} which allowed us to
define a Hopf algebra related to the Snyder
model \eqref{Snyder_orig1} - \eqref{Mx}. 
Then we have calculated the cross-commutation relations mimicking the Heisenberg double construction. However, in that framework it was not possible to define the dual elements to the Lorentz generators and the proper treatment of algebra of momenta $\tilde{A}$ would require using the quasi-Hopf algebra framework, hence we were not able to obtain the full Heisenberg double for the Snyder model. The full  Heisenberg double construction for the Snyder space within the Hopf algebra setting requires an introduction of the extended noncommutative coordinates which we discuss in this section.

Another way to obtain a Lie algebra from the Snyder model 
\eqref{Snyder_orig1} is to extend it by identifying the Snyder coordinates as $\hat{x}%
_{i}\sim \hat{x}_{iN}=M_{iN}\in so(1,N)/so(1,N-1)$ \cite{Girelli}, \cite
{2007Snyder}. This way one can define a Lie algebra corresponding to the extended
Snyder space and further one can also define the (coassociative) Hopf
algebra structure \cite{2007Snyder}. In this approach the Snyder coordinates
are seen as generators of the Lorentz algebra, but the Lorentz algebra
considered now has one dimension higher (i.e. $so(1,N)$ instead of $so(1,N-1)
$ from Sec. \ref{HD_Snyder}). Thanks to this unified extended \footnote{
Many authors use the word "\textit{generalized}" for the version of Snyder
space in a different meaning, see e.g. \cite{Ballb} or \cite{Tkachuk1}, 
therefore, following \cite{2007Snyder}, we shall call the version used here
as "\textit{extended}" instead of generalized - since it is unified with the additional tensorial coordinates transforming as the
Lorentz generators.} version of the Snyder model \cite{Girelli}, \cite
{2007Snyder} the Heisenberg double construction completely mimics the
construction of the undeformed Heisenberg double for the Lorentz algebra $%
so(1,N)$ with its dual algebra of functions on a group $SO(1,N)$ (Lorentz
matrices). In this way, the noncommutativity parameter $\beta$ related to the
Snyder space \eqref{Snyder_orig1} is implicitly included in the
cross-commutation relations. This will be considered in Section \ref
{HD_ext_lambdas}. In Section \ref{HD_ext_lambdas_real} we will also present the realizations for the
Lorentz matrices in the so-called Weyl realization \cite{2007Snyder}, \cite
{2003Melj}, \cite{Weyl_real}. However, to obtain the phase space from
the Heisenberg double we now can
consider a dual Hopf algebra of momenta and find the corresponding extended
Snyder phase space. In the remaining part of the paper, we want to focus on
finding the explicit formulae for such Heisenberg doubles corresponding to
the extended Snyder model.\\

We first need to define the Hopf algebra related to the
extended Snyder model and also define the dual Hopf algebra of objects which
would play the role of momenta. We start with embedding the Snyder algebra
relations \eqref{Snyder_orig1}, \eqref{MM}, \eqref{Mx}  in an algebra which
is generated by the $N$ position operators denoted by $\hat{x}_{i}$ and $%
N(N-1)/2$ antisymmetric tensorial coordinates $\hat{x}_{ij}$, transforming
as Lorentz generators \cite{Girelli}, \cite{2007Snyder}. This larger algebra
has the following commutation relations: 
\begin{eqnarray}
\lbrack \hat{x}_{i},\hat{x}_{j}] &=&i\lambda \beta \hat{x}_{ij},\qquad
\lbrack \hat{x}_{ij},\hat{x}_{kl}]=i\lambda (\eta _{ik}\hat{x}_{jl}-\eta_{il}%
\hat{x}_{jk}-\eta _{jk}\hat{x}_{il}+\eta _{jl}\hat{x}_{ik}),  \label{Snyder}
\\
\lbrack \hat{x}_{ij},\hat{x}_{k}] &=&i\lambda (\eta _{ik}\hat{x}%
_{j}-\eta_{jk}\hat{x}_{i}),  \label{Snyder2}
\end{eqnarray}%
where $\lambda $ and $\beta $ are real parameters. We can easily notice that
these commutation relations reduce to those of the standard Lorentz algebra
acting on commutative coordinates in the limit of $\beta \rightarrow 0$ and $%
\lambda \rightarrow 1$, and to the Lie algebra from Sec. \ref{HD_Snyder} (%
\eqref{Snyder_orig1}, \eqref{MM}, \eqref{Mx}) in the limit $\lambda
\rightarrow 1$.

To define the algebra in an unified way one can exploit the isomorphism between the Snyder
coordinates and the Lorentz generators of $so(1,N)$, and write the previous
formulas (\ref{Snyder}), (\ref{Snyder2}) more compactly defining, for
positive $\beta $, 
\begin{equation}  \label{Snyder_iso}
\hat{x}_{i}=\sqrt{\beta }\hat{x}_{iN}.
\end{equation}
The extended Snyder algebra then takes the form \cite{Girelli}, \cite%
{2007Snyder} 
of the Lorentz algebra $so(1,N)$ (to be precise it is $U_{so(1,N)}[[\lambda]]
$\footnote{A topological extension of the corresponding enveloping algebra $U_{so(1,N)}$ into an algebra of formal power series $U_{so(1,N)}[[\lambda]]$ in the formal parameter $\lambda$ is required here. This provides the $\lambda$-adic topology (see, for example, Chapter 1.2.10 in \cite{Klimyk}).}), given by one set of commutation relations as 
\begin{equation}  \label{Snyder_unified}
\lbrack \hat{x}_{\mu \nu },\hat{x}_{\rho \sigma }]=i\lambda (\eta _{\mu \rho
}\hat{x}_{\nu \sigma }-\eta _{\nu \rho }\hat{x}_{\mu \sigma }+\eta _{\nu
\sigma }\hat{x}_{\mu \rho }-\eta _{\mu \sigma }\hat{x}_{\nu \rho }),
\end{equation}
with $\eta _{NN}=1$ and $\eta _{kN}=0$, here $\mu =0,1,\dots N$ (Greek
indices are running from $0$ up to $N$, whereas the Latin indices are $%
i,j=0,1,\ldots, N-1$ as before).

One can check explicitly that (\ref{Snyder_unified}), via (\ref{Snyder_iso}),
reduces:

\begin{itemize}
\item to the Snyder noncommutative spacetime relations (\ref{Snyder}):
\begin{equation*}
[\hat{x}_{jN},\hat{x}_{iN}]=[\frac{1}{\sqrt{\beta }}\hat{x}_{j},\frac{1}{ 
\sqrt{\beta }}\hat{x}_{i}]=i\lambda (\eta _{ji}\hat{x}_{NN}-\eta _{Ni}\hat{x}
_{jN}+\eta _{NN}\hat{x}_{ji}-\eta _{jN}\hat{x}_{Ni})=i\lambda \hat{x}_{ji}
\end{equation*}
(note that $\hat{x}_{NN}=0$ due to antisymmetricity),
\item to the commutation relations for Lorentz generators (\ref{Snyder}):
\begin{equation*}
[ \hat{x}_{ij},\hat{x}_{kl}]=i\lambda (\eta _{ik}\hat{x}_{jl}-\eta_{il}\hat{x%
}_{jk}-\eta _{jk}\hat{x}_{il}+\eta _{jl}\hat{x}_{ik}),
\end{equation*}

\item and to cross-commutation relations of Lorentz generators acting on
coordinates (\ref{Snyder2}):
\begin{equation*}
[\hat{x}_{jk},\hat{x}_{iN}]=[\hat{x}_{jk},\frac{1}{\sqrt{\beta }}\hat{x}
_{i}]= i\lambda (\eta _{ji}\hat{x}_{kN}-\eta _{ki}\hat{x}_{jN}+\eta _{kN} 
\hat{x}_{ji}-\eta _{jN}\hat{x}_{ki})=
\end{equation*}
\begin{equation*}
=i\lambda \frac{1}{\sqrt{\beta }}(\eta _{ji} \hat{x}_{k}-\eta _{ki}\hat{x}
_{j}).
\end{equation*}
\end{itemize}

In turn, these all reduce to \eqref{Snyder_orig1}, \eqref{MM}, \eqref{Mx}
from Sec. \ref{HD_Snyder} respectively, for $\lambda \rightarrow 1$.

\subsection{Generalized Heisenberg algebra}

\label{gen_Heis} To discuss the extended phase space associated with this
extended Snyder model \eqref{Snyder_unified} as a result of the Heisenberg
double construction, we need to first recall few facts about the generalized
Heisenberg algebra and Weyl realization of the Lorentz algebra based on
results presented in \cite{2003Melj}.

The generalized Heisenberg algebra can be introduced as an unital,
associative algebra generated by (commutative) $x_{\mu \nu }$ and $p_{\mu
\nu }$ (both antisymmetric), satisfying the following commutation relations: 
\begin{eqnarray}  \label{genHeis}
\left[ x_{\mu \nu },x_{\alpha \beta }\right] &=&0, \\
\left[ p_{\mu \nu },p_{\alpha \beta }\right] &=&0,  \label{pp} \\
\left[ p_{\mu \nu },x_{\rho \sigma }\right] &=&-i\left( \eta _{\mu \rho
}\eta _{\nu \sigma }-\eta _{\mu \sigma }\eta _{\nu \rho }\right).  \label{px}
\end{eqnarray}
Here we consider the elements $p_{\mu \nu }$ as canonically conjugate to $%
x_{\mu \nu }$ which can be realized in standard way as $p_{\mu \nu }=-i\frac{%
\partial}{\partial x^{\mu\nu}}$.

Commutative coordinates $x_{\mu \nu }$ can be viewed as the classical limit
(when $\lambda \rightarrow 0$) of $\hat{x}_{\mu \nu }$ generators of $so(1,N)
$ used in the extended version for the Snyder algebra \eqref{Snyder_unified}%
. In other words, the Lie algebra $so(1,N)$, more specifically its universal
enveloping algebra $U_{so(1,N)}[[\lambda]]$, generated by $\hat{x}_{\mu \nu }
$ can be seen as a deformation of the underlying commutative space $x_{\mu
\nu }$ with $\lambda$ as the deformation parameter.

Now, since we are interested in the Snyder model (in its extended version)
and the corresponding deformed phase space, first let us notice that we can
make use of the analogous relation to \eqref{Snyder_iso} for the commutative
coordinates, i.e. take $x_{i}=\sqrt{\beta }x_{iN}$ and similarly for the
conjugate canonical momenta $p_{\mu \nu }$ we can introduce: $p_{i}=\frac{%
p_{iN}}{\sqrt{\beta }}$. This allows us to reduce the above generalized
Heisenberg algebra (\ref{genHeis}) - \eqref{px} to:

\begin{itemize}
\item the usual Heisenberg algebra sector, i.e. quantum mechanical phase
space corresponding to the commutative (classical) space-time: 
\begin{eqnarray}  \label{QMph-sp1}
\left[ x_{i},x_{j}\right] &=&0,\quad \left[ p_{i},p_{j}\right] =0 \\
\left[ p_{j},x_{i}\right] &=&\left[ p_{iN},x_{jN}\right] =-i\left( \eta
_{ij}\eta _{NN}-\eta _{iN}\eta _{Nj}\right) =-i\eta _{ij},  \label{QMph-sp2}
\end{eqnarray}

\item and "the remaining part", consisting of commutation relations between $%
x_{ij}$ - tensorial coordinates and $p_{ij}$ - their corresponding canonical
momenta: 
\begin{eqnarray}
\left[ x_{ij},x_{kl}\right] &=&\left[ x_{ij},x_{k}\right] =0,\quad \\
\left[ p_{ij},p_{kl}\right] &=&\left[ p_{ij},p_{k}\right] =0, \\
\left[ p_{ij},x_{kl}\right] &=&-i\left( \eta _{ik}\eta _{jl}-\eta _{il}\eta
_{jk}\right) , \\
\left[ p_{i},x_{kl}\right] &=&\left[ p_{ij},x_{k}\right] =0.
\end{eqnarray}
\end{itemize}

Therefore, the relations (\ref{genHeis})-\eqref{px} indeed describe the
generalization of the quantum mechanical phase space as they contain, as a
subalgebra, the relations (\ref{QMph-sp1})-\eqref{QMph-sp2}. 

The Weyl realization of the Lorentz algebra in terms of this generalized
Heisenberg algebra (\ref{genHeis})-\eqref{px} (as formal power series) have
been discussed in detail in \cite{2003Melj}.

\section{Extended Snyder space and its Heisenberg doubles}\label{extSn-HDs}

We are now ready to discuss how to construct two full Heisenberg doubles
corresponding to the extended Snyder space \eqref{Snyder_unified}, in its
unified $so(1,N)$-like version (with the generators $\hat{x}_{\mu \nu}$). 
For the purpose of this section let us denote the extended Snyder algebra,
defined by relations \eqref{Snyder_unified} as an algebra $B$. To construct
the Heisenberg double for the extended Snyder algebra $B$ 
we first need to equip it in the Hopf algebra structure (which is
straightforward as we can use the primitive coproducts for $\hat{x}_{\mu \nu
}$) and then we need to define the dual Hopf algebra.

We equip the Snyder algebra $B$ (as $so(1,N)$), defined by relations (\ref%
{Snyder_unified}), with the Hopf algebra structure as follows: 
\begin{equation}  \label{copx}
\Delta \left( \hat{x}_{\mu \nu }\right) =\Delta _{0}\left( \hat{x}_{\mu \nu
}\right) ,
\end{equation}%
\begin{equation}  \label{Sx}
\epsilon \left( \hat{x}_{\mu \nu }\right) =0\mbox{\ and\ }S\left( \hat{x}%
_{\mu \nu }\right) =-\hat{x}_{\mu \nu }.
\end{equation}
With the above relations algebra $B$, using \eqref{Snyder_iso}, leads to the
extended Snyder Hopf algebra.

\subsection{Extended Snyder phase space from the Heisenberg double
construction}
\label{HD_ext_momenta}

To discuss the phase space corresponding to the extended Snyder space, we consider the Hopf algebra generated by $p_{\mu
\nu }$ (satisfying (\ref{pp})), as a
dual Hopf algebra $B^*$, which is equipped with the Hopf algebra
structure introduced in \cite{2007Snyder}
[see, eq.(25) therein] where the coproducts, calculated up to the third
order, have the following form: 
\begin{eqnarray}  \label{copp}
\Delta p_{\mu \nu } &=&\Delta _{0}p_{\mu \nu }-\frac{\lambda }{2}\left(
p_{\mu \alpha }\otimes p_{\nu \alpha }-p_{\nu \alpha }\otimes p_{\mu \alpha
}\right) +  \notag \\
&&-\frac{\lambda ^{2}}{12}( p_{\mu \alpha }\otimes p_{\alpha \beta }p_{\nu
\beta }-p_{\nu \alpha }\otimes p_{\alpha \beta }p_{\mu \beta }-2p_{\alpha
\beta }\otimes p_{\mu \alpha }p_{\nu \beta }  \notag \\
&&+p_{\mu \alpha }p_{\alpha \beta }\otimes p_{\nu \beta }-p_{\nu \alpha
}p_{\alpha \beta }\otimes p_{\mu \beta }-2p_{\mu \alpha }p_{\nu \beta
}\otimes p_{\alpha \beta })+O(\lambda^3).
\end{eqnarray}
Counits are $\epsilon ( p_{\mu \nu }) =0$ and antipodes are $S( p_{\mu \nu }) =-
p_{\mu\nu}$. This defines the coassociative\footnote{%
In general, if noncommutative coordinates close a Lie algebra, as it is the
case for $\hat{x}_{\mu\nu}$, then the corresponding coproducts of momenta
are coassociative \cite{Lukierski_associative}, \cite{Weyl_real}.} Hopf algebra $B^*$ as the dual
to the extended Snyder Hopf algebra $B$. The
above coproducts for momenta are corresponding to the so-called Weyl
realization for the $\hat{x}_{\mu\nu}$\footnote{%
The Weyl realization for the extended Snyder space is defined as $e^{ik_i%
\hat{x}_i+\frac{i}{2}k_{ij}\hat{x}_{ij}}\triangleright 1 = e^{ik_ix_i+\frac{i%
}{2}k_{ij}x_{ij}}$ where $\hat{x}_{i}\triangleright 1={x}_{i},\quad\hat{x}%
_{ij}\triangleright 1={x}_{ij}$, see eq. (17) in \cite{2007Snyder}. Note
that the action differs from the one described in footnote 2.}. One could
use the coproducts for generic realization but they depend on 5 free
parameters and are calculated up to the second order in $\lambda$ \cite%
{2007Snyder}. 

The duality relation $<\quad ,\quad >:B^*\times B\rightarrow \mathbb{C}$ is
as follows: 
\begin{equation}  \label{duality}
<p_{\mu \nu },\hat{x}_{\rho \sigma }>=-i\left( \eta _{\rho \mu }\eta
_{\sigma \nu }-\eta _{\sigma \mu }\eta _{\rho \nu }\right)
\end{equation}
and can be extended to all elements of both Hopf algebras $B$ and $B^*$.
We take the left Hopf action $\triangleright $ 
of $B^*$ on $B$, which is defined by 
\begin{equation}
p_{\rho \sigma }\triangleright \hat{x}_{\mu \nu }=<p _{\rho \sigma },\hat{x}
_{\mu \nu _{\left( 2\right) }}>\hat{x}_{\mu \nu _{\left( 1\right)
}}=-i\left( \eta _{\rho \mu }\eta _{\sigma \nu }-\eta _{\sigma \mu }\eta
_{\rho \nu }\right).
\end{equation}
We can now construct the corresponding Heisenberg double resulting in the
following cross-commutation relations: 
\begin{eqnarray}  \label{ext_ph_sp}
\left[ p_{\mu \nu },\hat{x}_{\rho \sigma }\right] &=&\hat{x}_{\rho \sigma
}{}_{\left( 1\right) }<p_{\mu \nu \left( 1\right) },\hat{x}_{\rho \sigma
\left( 2\right) }>p_{\mu \nu \left( 2\right) }-\hat{x}_{\rho \sigma }p_{\mu
\nu }=  \notag \\
&=&-i\left( \eta _{\rho \mu }\eta _{\sigma \nu }-\eta _{\sigma \mu }\eta
_{\rho \nu }\right) +\frac{i\lambda }{2}(\eta _{\rho \mu }p_{\nu \sigma
}-\eta _{\sigma \mu }p_{\nu \rho }-\eta _{\rho \nu }p_{\mu \sigma }+\eta
_{\sigma \nu }p_{\mu \rho })+  \notag \\
+\frac{i\lambda ^{2}}{12}[\eta _{\rho \mu }p_{\sigma \beta }p_{\nu \beta
}&-&\eta _{\sigma \mu }p_{\rho \beta }p_{\nu \beta }-\eta _{\rho \nu
}p_{\sigma \beta }p_{\mu \beta }+\eta _{\sigma \nu }p_{\rho \beta }p_{\mu
\beta }-2p_{\mu \rho }p_{\nu \sigma }+2p_{\mu \sigma }p_{\nu \rho }]+O\left(
\lambda ^{3}\right).
\end{eqnarray}
For the description of the phase space corresponding to the extended Snyder
model (in Snyder coordinates $\hat{x_i}$, $\hat{x}_{ij}$), we make use of
the isomorphism (\ref{Snyder_iso}) $\hat{x}_{i}= \sqrt{\beta }\hat{x}_{iN}$
and $p_{i}=\frac{p_{iN}}{\sqrt{\beta }}.$ The above duality (\ref{duality})
then becomes: 
\begin{eqnarray}
&&<p_{j},\hat{x}_{i}>=-i\eta _{ij},  \label{dual_red1} \\
&&<p_{k},\hat{x}_{ij}>=0,  \label{dual_red2} \\
&&<p_{kl},\hat{x}_{i}>=0, \\
&& <p_{kl},\hat{x}_{ij}>=-i\left( \eta _{ik}\eta _{jl}-\eta _{jk}\eta
_{il}\right).
\end{eqnarray} 
The cross-commutation relations are as follows:
\begin{itemize}
\item commutation relations between Snyder coordinates and their coupled momenta: 
\begin{eqnarray}
\left[ p_{k},\hat{x}_{i}\right] &=&\left[ p_{kN},\hat{x}_{iN}\right] =
-i\eta_{ik} (1-\frac{\beta\lambda ^{2}}{12} p_{l}p_{l})-\frac{i\beta\lambda
^{2}}{12} p_{k}p_{i} +\frac{i\lambda }{2}p_{ki}+\frac{i\lambda ^{2}}{12}%
p_{il}p_{kl}+O\left( \lambda ^{3}\right),  \label{our_ph_sp1}
\end{eqnarray}
\item commutation relations between tensorial coordinates and their coupled
momenta: 
\begin{eqnarray}
\left[ p_{kl},\hat{x}_{ij}\right] &=&-i\left( \eta _{ik}\eta _{jl}-\eta
_{jk}\eta _{il}\right) +i\frac{\lambda }{2}(\eta _{ik}p_{lj}-\eta
_{jk}p_{li}-\eta _{il}p_{kj}+\eta _{jl}p_{ki})+\frac{i\lambda ^{2}}{12}[(
\eta _{ik}p_{jm}p_{lm}-\eta _{jk}p_{im}p_{lm})  \notag \\
&-&\left( \eta _{il}p_{jm}p_{km}-\eta _{jl}p_{im}p_{km}\right)
-2p_{ki}p_{lj}+2p_{kj}p_{li}]+O\left( \lambda ^{3}\right),  \label{our_ph_sp2}
\end{eqnarray}
\item and mixed relations: 
\begin{equation}
\left[ p_{kl},\hat{x}_{i}\right] =\left[ p_{kl},\sqrt{\beta }\hat{x}_{iN}%
\right] =i\frac{\lambda }{2}\beta (\eta _{ik}p_{l}-\eta _{il}p_{k})-i\frac{%
\lambda ^{2}}{12}\beta \lbrack \eta _{ik}p_{m}p_{lm}-\eta
_{il}p_{m}p_{km}+2p_{ki}p_{l}-2p_{k}p_{li}]+O\left( \lambda ^{3}\right),
\label{our_ph_sp3}
\end{equation}
\begin{equation}
\left[ p_{k},\hat{x}_{ij}\right] =\left[ \frac{1}{\sqrt{\beta }}p_{kN},\hat{x%
}_{ij}\right] =-i\frac{\lambda }{2}(\eta _{ik}p_{j}-\eta _{jk}p_{i})-i\frac{%
\lambda ^{2}}{12}[\left( \eta _{ik}p_{jl}p_{l}-\eta _{jk}p_{il}p_{l}\right)
-2p_{ki}p_{j}+2p_{kj}p_{i}]+O\left( \lambda ^{3}\right).  \label{our_ph_sp4}
\end{equation}
\end{itemize}
One can notice that the commutator between the momenta generators and the Snyder
coordinates \eqref{our_ph_sp1} obtained from \eqref{ext_ph_sp} actually
resembles the Snyder model phase space for the Weyl realization %
\eqref{ph_spWeyl} obtained in Sec.\ref{HD_Snyder}. The first three terms
agree up to the factor $\frac{\lambda^2}{4}$ but the remaining terms include
the tensorial momenta.

We have now obtained the full extended Snyder phase space \eqref{our_ph_sp1}-%
\eqref{our_ph_sp4} 
resulting from the Heisenberg double construction.

It is worth to mention that some authors, see e.g. \cite%
{Balla}, also consider another version of the Snyder phase spaces, where
momenta do not commute, 
but we have not considered
this type of phase spaces in this work, the momenta sector is always commutative,
for both the Snyder model considered in Sec. \ref{HD_Snyder} and the
extended Snyder model in Sec. \ref{HD_ext_momenta}.

\subsection{Another Heisenberg double for the extended Snyder algebra}

\label{HD_ext_lambdas} To construct another Heisenberg double for the
extended Snyder algebra $B$ written in an unified $so(1,N)$-like form %
\eqref{Snyder_unified}, it is quite straightforward to mimic the Heisenberg
double construction for the Lorentz algebra.

We take the extended Snyder Hopf algebra $B$ \eqref{Snyder_unified} 
equipped with the Hopf algebra structure (\ref{copx}),(\ref{Sx}), as before. 
We define the dual Hopf algebra algebra $D$ (as the dual to the extended
Snyder Hopf algebra $B$) as an algebra of functions on a group $SO(1,N)~$%
which is generated by Lorentz matrices $\Lambda _{\alpha \beta }$, i.e.: %
\begin{eqnarray}  \label{Ddual}
D &=&F\left( SO(1,N)\right) =\{\Lambda _{\alpha \beta }:\left[ \Lambda
_{\alpha \beta },\Lambda _{\mu \nu }\right] =0:\Lambda ^{T}\eta \Lambda
=\eta \}, \\
\Delta \left( \Lambda _{\rho \sigma }\right) &=&\Lambda _{\rho \alpha
}\otimes \Lambda _{\alpha \sigma };\quad \epsilon \left( \Lambda _{\rho
\sigma }\right) =\delta _{\rho \sigma }\quad ;S\left( \Lambda _{\rho \sigma
}\right) =(\Lambda ^{-1})_{\rho \sigma }=\Lambda _{\sigma \rho }.
\end{eqnarray}
Note that the Greek indices are running up to $N$, i.e. $\alpha,\beta=0,1,\ldots, N$.
The duality relation is $<\quad ,\quad >:D\times B\rightarrow \mathbb{C}$ is
given by: 
\begin{equation}
<\Lambda _{\rho \sigma },\hat{x}_{\mu \nu }>=-i\lambda (\eta _{\rho \mu
}\eta _{\sigma \nu }-\eta _{\rho \nu }\eta _{\sigma \mu }).  \label{dual}
\end{equation}
We consider the left Hopf action $\triangleright$ 
of $D$ on $B$, which is defined by 
\begin{equation}
\Lambda _{\rho \sigma }\triangleright \hat{x}_{\mu \nu }=<\Lambda _{\rho
\sigma },\hat{x}_{\mu \nu _{\left( 2\right) }}>\hat{x}_{\mu \nu _{\left(
1\right) }}=\eta _{\rho \sigma }\hat{x}_{\mu \nu }-i\lambda (\eta _{\rho \mu
}\eta _{\sigma \nu }-\eta _{\rho \nu }\eta _{\sigma \mu }).
\end{equation}
And we calculate the cross-commutation relations defining the Heisenberg
double as: 
\begin{eqnarray}  \label{Lx}
\left[ \Lambda _{\rho \sigma },\hat{x}_{\mu \nu }\right] &=&\hat{x}_{\mu \nu
}{}_{\left( 1\right) }<\Lambda _{\rho \sigma \left( 1\right) },\hat{x}_{\mu
\nu \left( 2\right) }>\Lambda _{\rho \sigma \left( 2\right) }-\hat{x}_{\mu
\nu }\Lambda _{\rho \sigma } \\
&=&-i\lambda (\eta _{\rho \mu }\Lambda _{\nu \sigma }-\eta _{\rho \nu
}\Lambda _{\mu \sigma }).  \label{Lx}
\end{eqnarray}
For the description of the Heisenberg double for the extended Snyder model
(in Snyder coordinates $\hat{x}_i$, $\hat{x}_{ij}$), we again make use of
the isomorphism \eqref{Snyder_iso}, 
and the above formulae lead to:
\begin{itemize}
\item duality for the {Snyder coordinates} with Lorentz matrices: 
\begin{eqnarray}
&&<\Lambda _{jk},\hat{x}_{i}>=\sqrt{\beta }<\Lambda _{jk},\hat{x}%
_{iN}>=-i\lambda \sqrt{\beta }(\eta _{ji}\eta _{kN}-\eta _{jN}\eta _{ki})=0,
\label{dualLx3} \\
&&<\Lambda _{jN},\hat{x}_{i}>=\sqrt{\beta }<\Lambda _{jN},\hat{x}%
_{iN}>=-i\lambda \sqrt{\beta }\eta _{ji}, \\
&&<\Lambda _{Nk},\hat{x}_{i}>=\sqrt{\beta }<\Lambda _{Nk},\hat{x}%
_{iN}>=i\lambda \sqrt{\beta }\eta _{ki}, \\
&&<\Lambda _{NN},\hat{x}_{i}>=\sqrt{\beta }<\Lambda _{NN},\hat{x}_{iN}>=0,
\end{eqnarray}
\item duality of the Lorentz generators with their dual Lorentz matrices: 
\begin{eqnarray}
&&<\Lambda _{jk},\hat{x}_{ip}>=-i\lambda (\eta _{ji}\eta _{kp}-\eta
_{jp}\eta _{ki}),  \label{dualLM3} \\
&&<\Lambda _{jN},\hat{x}_{ip}>=0=<\Lambda _{Nk},\hat{x}_{ip}>, \\
&&<\Lambda _{NN},\hat{x}_{ip}>=0.
\end{eqnarray}
\end{itemize}
Similarly the cross-commutation relations from the Heisenberg double construction (\ref{Lx})
then become as follows:
\begin{itemize}
\item cross-commutation relations between the {Snyder coordinates} and
Lorentz matrices: 
\begin{eqnarray}
\left[ \Lambda _{jk},\hat{x}_{i}\right] &=&\sqrt{\beta }\left[ \Lambda _{jk},%
\hat{x}_{iN}\right] =-i\lambda \sqrt{\beta }\eta _{ji}\Lambda _{Nk}, \\
\left[ \Lambda _{jN},\hat{x}_{i}\right] &=&\sqrt{\beta }\left[ \Lambda _{jN},%
\hat{x}_{iN}\right] =-i\lambda \sqrt{\beta }\eta _{ji}\Lambda _{NN}, \\
\left[ \Lambda _{Nk},\hat{x}_{i}\right] &=&\sqrt{\beta }\left[ \Lambda _{Nk},%
\hat{x}_{iN}\right] =i\lambda \sqrt{\beta }\Lambda _{ik}, \\
\left[ \Lambda _{NN},\hat{x}_{i}\right] &=&\sqrt{\beta }\left[ \Lambda _{NN},%
\hat{x}_{iN}\right] =i\lambda \sqrt{\beta }\Lambda _{iN},
\end{eqnarray}
\item and the cross-commutation relations between the Lorentz generators (of 
$so(1,N-1)$) with the Lorentz matrices: 
\begin{eqnarray}
\left[ \Lambda _{jk},\hat{x}_{ip}\right] &=&-i\lambda (\eta _{ji}\Lambda
_{pk}-\eta _{jp}\Lambda _{ik}), \\
\left[ \Lambda _{jN},\hat{x}_{ip}\right] &=&-i\lambda (\eta _{ji}\Lambda
_{pN}-\eta _{jp}\Lambda _{iN}), \\
\left[ \Lambda _{Nk},\hat{x}_{ip}\right] &=&-i\lambda (\eta _{Ni}\Lambda
_{pk}-\eta _{Np}\Lambda _{ik})=0, \\
\left[ \Lambda _{NN},\hat{x}_{ip}\right] &=&0.
\end{eqnarray}
\end{itemize}
The primitive coproduct for $\hat{x}_{\mu \nu }$ reduces to the primitive
coproduct for $\hat{x}_{ip}$ (Lorentz generators of $so(1,N-1)$) and to the
primitive coproduct for $\hat{x}_{i}$ (Snyder coordinates), as in Sec. \ref{HD_Snyder}, respectively.

We also note that 
\begin{equation}
\lbrack \Lambda _{\mu \nu },p_{\rho \sigma }]=0.
\end{equation}

\subsection{Realizations for Lorentz matrices}\label{HD_ext_lambdas_real}
We can actually relate the dual momenta discussed in Sec. \ref{HD_ext_momenta} with the dual Lorentz matrices discussed in  Sec. \ref{HD_ext_lambdas}.

This can be done by introducing the realizations of the elements of the dual
algebra $D $, i.e. functions on a group $SO(1,N)$ - Lorentz matrices $%
\Lambda_{\alpha\beta}$. 
These realizations can be expressed as a formal power series of the
tensorial momenta introduced in Sec. \ref{gen_Heis}. For more details we
refer the reader to \cite{2003Melj} where the Lorentz algebra extension by
its dual counterpart has been discussed in detail. The formulas presented
below base on the Theorem III.1 from \cite{2003Melj}. The Weyl realization 
\cite{2003Melj}, \cite{Weyl_real} for the generators of the algebra $D$,
satisfying \eqref{Ddual}, can be written in a very compact form as follows: 
\begin{equation}
\Lambda _{\rho \sigma }=\left( e^{\lambda p}\right) _{\rho \sigma }.
\end{equation}
If we want to calculate the explicit formulae for the realization of the
elements dual to the Lorentz sector in the extended Snyder algebra written
in the form of tensorial coordinates $\hat{x}_{ij}$ \eqref{Snyder}, %
\eqref{Snyder2} we use the above formula and obtain: 
\begin{eqnarray}
\Lambda _{kl} &=&\sum_{n=0}^{\infty }\frac{\lambda ^{n}}{n!}\left(
p^{n}\right) _{kl},  \notag \\
\Lambda _{Nl} &=&\sum_{n=1}^{\infty }\frac{\lambda ^{n}}{n!}p_{Nk}\left(
p^{n-1}\right) _{kl}=p_{Nk}\left( \frac{\Lambda -\eta }{p}\right) _{kl}, 
\notag \\
\Lambda _{kN} &=&\sum_{n=0}^{\infty }\frac{\lambda ^{n}}{n!}\left(
p^{n}\right) _{kN}=\sum_{n=1}^{\infty }\frac{\lambda ^{n}}{n!}\left(
p^{n-1}\right) _{kl}p_{lN}=\left( \frac{\Lambda -\eta }{p}\right)
_{kl}p_{lN}=\left( \Lambda ^{-1}\right) _{Nk},  \notag \\
\Lambda _{NN} &=&\sum_{n=0}^{\infty }\frac{\lambda ^{n}}{n!}\left(
p^{n}\right) _{NN}=\eta _{NN}+\sum_{n=2}^{\infty }\frac{\lambda ^{n}}{n!}%
p_{Nk}\left( p^{n-2}\right) _{kl}p_{lN}=\eta _{NN}+p_{Nk}\left( \frac{%
\Lambda -\eta -\lambda p}{p^2}\right) _{kl}p_{lN},  \notag
\end{eqnarray}
where we also used the following notation $(p^{0})_{\alpha \beta }=\eta _{\alpha
\beta }$ and $\left( \Lambda ^{0}\right) _{\rho \sigma }=\eta _{\rho \sigma
} $.

\section{Conclusions}

In this paper we have investigated three Heisenberg doubles related to the
two types of noncommutative Snyder models. The Heisenberg double
construction was widely investigated for other noncommutative spacetimes 
\cite{KosMas}, \cite{GGKM}, \cite{Lukierski1}, \cite{Lukierski2}, \cite{TMPH}%
, but not yet (up to our knowledge) for the Snyder space. Therefore this
work offers the first study on Heisenberg doubles for the Snyder model as
well as for the extended Snyder model.

In Sec. \ref{HD_Snyder} we discuss issues arising when applying the Heisenberg double construction to the Snyder model. We propose a
duality between Snyder coordinates and momenta with the (non-coassociative) coproducts related to the so-called Snyder realization \eqref{copp_Sn}. This duality is valid on the generators (in the linear powers) only.
The cross-commutation relations obtained for the generators \eqref{p-x1} are compared with the Snyder phase space relations considered in the literature.
Then we use the momenta with the (non-coassociative) coproducts  in the general realization which provides more general version of the cross-commutation relations between the momenta and Snyder coordinates and reduces to all known cases for the certain choices of the
parameter $c$ (which parametrizes the non-coassociative coproduct for the
momenta generators). However, at the end of Sec. \ref{HD_Snyder}, we point out that the construction of the full  Heisenberg double for the Snyder space in the Hopf algebra setting requires the introduction of extended noncommutative coordinates.

Therefore, we use the fact that the Snyder model can be embedded in a larger
algebra: $U_{so(1,N)}[[\lambda]]$, for which the dual algebra admits the coassociative
coalgebra structure. We then construct the Heisenberg double for this
extended Snyder model in two ways. Firstly, by introducing the dual
tensorial momentum space. Secondly, by using the Lorentz matrices, i.e.
functions on the Lorentz group. 

The Heisenberg double of the extended Snyder
algebra with their dual momenta can be interpreted as the extended Snyder
phase space. The formulation of Heisenberg doubles as extended Snyder phase
spaces, proposed in this work, can be used in many further applications. Additional
advantage is that, since the noncommutative coordinates generate the Lie
algebra then the corresponding coproducts of momenta are coassociative, and
the related star products between coordinates are associative \cite%
{2007Snyder}, which opens a way to a great number of applications where the
associative star product is required. The only drawback of this approach is
the unclear physical interpretation of tensorial coordinates and
corresponding conjugated momenta appearing in this picture. 

Nevertheless, by using the algebraic scheme of Heisenberg doubles, one can
introduce the covariant Snyder phase spaces and further investigate the applications where the consistent
definition of a phase space is crucial, for example such deformed (extended)
Snyder phase space may lead to deformed Heisenberg uncertainty relations 
\cite{PRD79}, or may be considered in the context of quantum gravity
phenomenology \cite{Battisti:2008du}, \cite{Mig2}, \cite{Mig3}, or when
investigating cosmological \cite{Battisti:2008du} and curved \cite{Mig1a}, \cite{Mig1b} backgrounds coupled to Snyder spacetime. We
consider our work presented here as a first step towards such applications.

\section*{Acknowledgements}
The authors would like to thank J. Lukierski and Z. \v Skoda for useful comments.
AP would like to acknowledge the contribution of the COST Action CA18108.
\section*{Appendix 1: Heisenberg double construction.}

Let $A$ and $A^*$ be dual Hopf algebras. To construct the Heisenberg double $%
A\rtimes A^{\ast }$ we start with the left Hopf action $\triangleright $ of $%
A^{\ast }$ on $A$ defined as: 
\begin{equation}
a^{\ast }\triangleright a=<a^{\ast },a_{\left( 2\right) }>a_{(1)},
\end{equation}%
where we use the Sweedler notation $\Delta \left( a\right) =a_{\left(
1\right) }\otimes a_{\left( 2\right) }$ for the coproduct and $a^{\ast }\in
A^{\ast },a\in A$.

Duality needs to satisfy the following compatibility conditions between
algebras: 
\begin{eqnarray}
<a_{\left( 1\right) }^{\ast },a><a_{\left( 2\right) }^{\ast },a^{\prime
}>&=&<a^{\ast },a\cdot a^{\prime }>, \\
<a^{\ast },a_{\left( 1\right) }><a^{\ast \prime },a_{\left( 2\right)
}>&=&<a^{\ast }\cdot a^{\ast \prime },a>.
\end{eqnarray}
The Heisenberg double corresponding to these data can be then constructed as
the crossed product algebra (aka \textquotedblleft smash
product\textquotedblright ) $A\rtimes A^{\ast }.$ The (left) product in the
crossed product algebra (Heisenberg double) becomes: 
\begin{equation}
(a\otimes a^{\ast })\rtimes (a^{\prime }\otimes a^{\ast \prime })=a\left(
a_{\left( 1\right) }^{\ast }\triangleright a^{\prime }\right) \otimes
a_{\left( 2\right) }^{\ast }a^{\ast \prime }=<a_{\left( 1\right) }^{\ast
},a_{\left( 2\right) }^{\prime }>aa^{\prime }{}_{\left( 1\right) }\otimes
a_{\left( 2\right) }^{\ast }a^{\prime },
\end{equation}
which leads to the following (left) products:%
\begin{eqnarray}
a^{\ast }\circ a &=&(1\otimes a^{\ast })\rtimes (a\otimes 1)=\left(
a_{\left( 1\right) }^{\ast }\triangleright a\right) \otimes a_{\left(
2\right) }^{\ast }=<a_{\left( 1\right) }^{\ast },a_{\left( 2\right)
}>a{}_{\left( 1\right) }\otimes a_{\left( 2\right) }^{\ast }, \\
a\circ a^{\ast } &=&(a\otimes 1)\rtimes (1\otimes a^{\ast }).
\end{eqnarray}
So the cross-commutation relation becomes: 
\begin{equation}
\left[ a^{\ast },a\right] =a{}_{\left( 1\right) }<a_{\left( 1\right) }^{\ast
},a_{\left( 2\right) }>a_{\left( 2\right) }^{\ast }-a\circ a^{\ast }.
\label{left_cop}
\end{equation}

\bigskip

There are some special cases worth considering:

\begin{enumerate}
\item If the generators of $A$ have primitive coproduct then the above formula, for the generators, reduces
to: 
\begin{equation}
\left[ a^{\ast },a\right] =<a_{\left( 1\right) }^{\ast },a>a_{\left(
2\right) }^{\ast }.
\end{equation}

\item If the coproduct on the algebra $A^{\ast }$ is opposite, i.e. $\Delta
a^{\ast }=a_{\left( 2\right) }^{\ast }\otimes a_{\left( 1\right) }^{\ast }$
then the commutator becomes: 
\begin{equation}
\left[ a^{\ast },a\right] =a{}_{\left( 1\right) }<a_{\left( 2\right) }^{\ast
},a_{\left( 2\right) }>a_{\left( 1\right) }^{\ast }-a\circ a^{\ast }.
\end{equation}
And if, in addition the generators of $A$ have primitive coproduct then,for generators, we get: 
\begin{equation}
\left[ a^{\ast },a\right] =<a_{\left( 2\right) }^{\ast },a>a_{\left(
1\right) }^{\ast }.
\end{equation}
\end{enumerate}

\end{document}